\documentclass[twocolumn,aps,pra,longbibliography,floatfix,superscriptaddress,preprintnumbers,10pt]{revtex4-1}
\usepackage{color}
\usepackage{amsmath}
\usepackage{amssymb}
\usepackage{graphicx}
\makeatletter
\usepackage{times}

\usepackage[hypertexnames=false,naturalnames=true]{hyperref}
\hypersetup{colorlinks=True}
\usepackage[all]{hypcap}

\begin{document}

\title{Dissipation without resistance: Imaging impurities at quantum Hall edges}

\author{Gu Zhang}
\affiliation{Institute for Quantum Materials and Technologies, Karlsruhe Institute of Technology, 76021 Karlsruhe, Germany}
\affiliation{Institut f\"ur Nanotechnologie, Karlsruhe Institute of Technology, 76021 Karlsruhe, Germany}
\author{Igor V. Gornyi}
\affiliation{Institute for Quantum Materials and Technologies, Karlsruhe Institute of Technology, 76021 Karlsruhe, Germany}
\affiliation{Institut f\"ur Nanotechnologie, Karlsruhe Institute of Technology, 76021 Karlsruhe, Germany}
\affiliation{A. F. Ioffe Physico-Technical Institute, 194021 St.~Petersburg, Russia}
\affiliation{\mbox{Institut f\"ur Theorie der Kondensierten Materie, Karlsruhe Institute of Technology, 76128 Karlsruhe, Germany}}
\author{Alexander D. Mirlin}
\affiliation{Institute for Quantum Materials and Technologies, Karlsruhe Institute of Technology, 76021 Karlsruhe, Germany}
\affiliation{Institut f\"ur Nanotechnologie, Karlsruhe Institute of Technology, 76021 Karlsruhe, Germany}
\affiliation{\mbox{Institut f\"ur Theorie der Kondensierten Materie, Karlsruhe Institute of Technology, 76128 Karlsruhe, Germany}}
\affiliation{Petersburg Nuclear Physics Institute, 188350 St.~Petersburg, Russia}
\affiliation{L.\,D.~Landau Institute for Theoretical Physics RAS, 119334 Moscow, Russia}
\date{\today}

\begin{abstract}

Motivated by the recent experiment by Marguerite \textit{et al.} \cite{Zeldov2019} on imaging in graphene samples, we investigate theoretically the dissipation induced by resonant impurities in the quantum Hall regime.
The impurity-induced forward scattering of electrons at quantum Hall edges leads to an enhanced phonon emission, which reaches its maximum when the impurity state is tuned to resonance by a scanning tip voltage.
Our analysis of the effect of the tip potential on the dissipation reveals peculiar thermal rings around the impurities, in consistency with experimental observations.
Remarkably, this impurity-induced dissipation reveals non-trivial features that are unique for chiral 1D systems such as quantum Hall edges.
First, the dissipation is not accompanied by the generation of resistance.
Second, this type of dissipation is highly nonlocal: a single impurity induces heat transfer to phonons along the whole edge.

\end{abstract}

\maketitle

\section{Introduction}

Although the quantum Hall (QH) effect has been studied for decades,  it continues to attract attention of the community for multiple reasons.
To begin with, at the QH plateaux, mobile quasiparticles are confined within the sample edges.
Consequently, QH problems bridge two-dimensional (2D) and one-dimensional (1D) physics, with the latter 
allowing for exact description in terms of bosonization and other well-developed techniques \cite{GiamarchiBook, FendleyLudwigSaleurPRL95, KaneFisherPRB95, QSHPRL05}.
The second reason is the chiral nature of edge excitations, which leads to the most distinguished feature of QH effect: the topologically protected quantized Hall conductance.
This feature is one of the major reasons of the long-lasting interest on QH effect.
Furthermore, multiple QH-based complex structures have been proposed as possible hosts of quasiparticles with exotic statistics \cite{AliceaMajoranaReview,Lindner12,ChengPRB12, FrankPRB13, ClarkeAliceaShtengelNatPhys14, Mong14, SanJose15,AliceaFendleyAnnualReview16}.

Conventionally, dissipation in electronic transport is related to resistance.
One could thus expect that the absence of backscattering in the topologically protected edge state is accompanied by the absence of energy dissipation.
However, it has been shown that even for a clean single chiral channel, energy dissipation is possible through the electron-phonon interaction \cite{SlizovskiyFalkoPRB17}.
This effect leads to dissipation (transfer of energy from electrons to phonons) along the quantum Hall edges with a homogeneous power density. It can be considered as an ultimate manifestation of "nonlocal dissipation" \cite{RokniLevinson1995,Tikhonov2019}. Indeed, the work on the electronic system is performed only at the contacts between the QH edge and reservoirs, while the energy transfer to phonons is spatially separated from the regions producing the electrical resistance.

\begin{figure}
\centering
\includegraphics[width = 0.45 \textwidth]{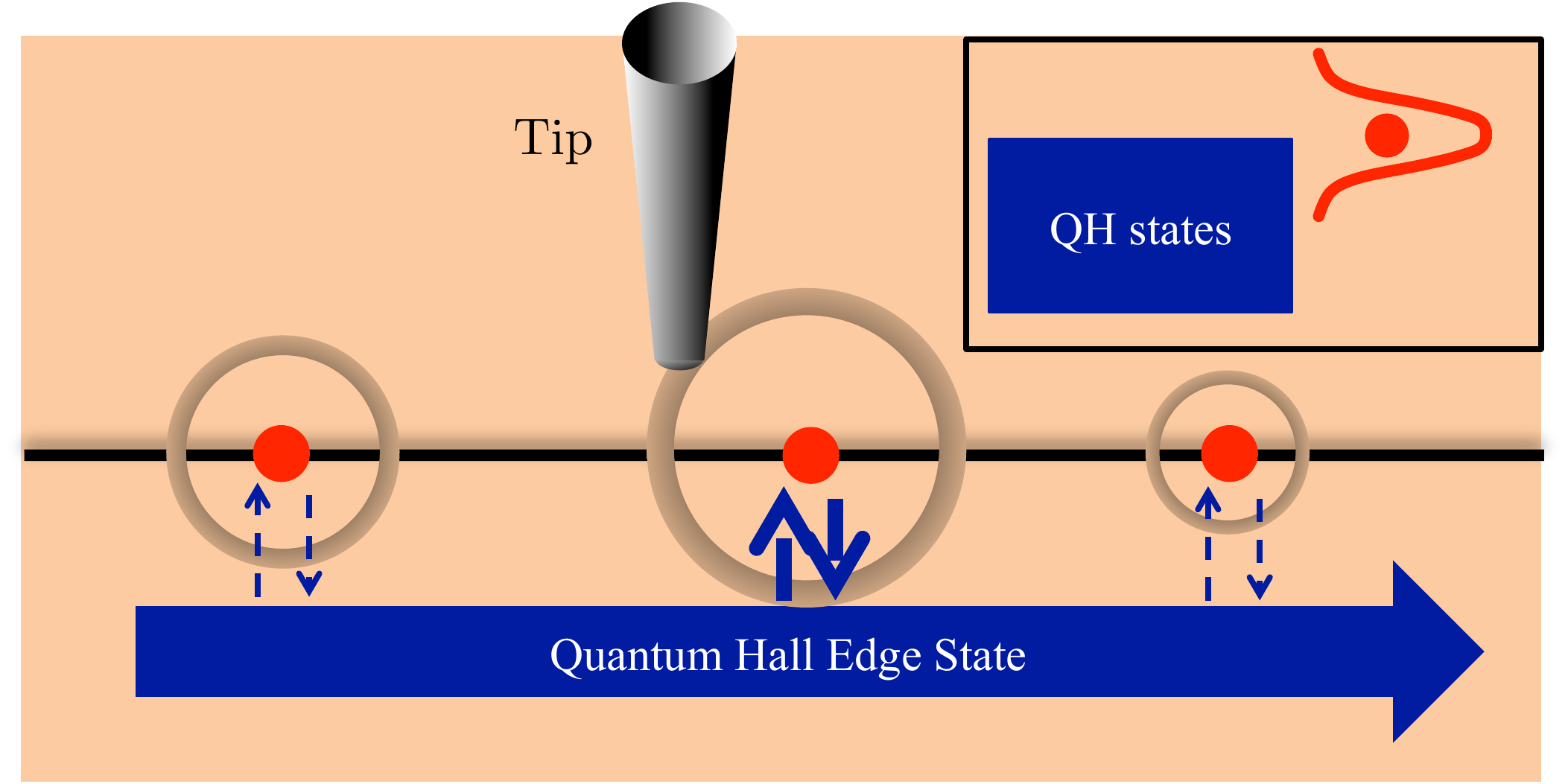}
\caption{The thermal rings induced by the impurity (red dot) at the sample boundary (thick black line). When the tip is placed on top of a ``thermal ring'', the tip voltage tunes the impurity to resonance, leading to a stronger impurity-QH tunneling and an enhanced energy dissipation rate. This leads to a higher temperature at the rings that is represented by the darker color. The corresponding impurity energy diagram  is shown in the inset.}
\label{fig:rings}
\end{figure}

At the same time, realistic graphene samples contain abundance of resonant impurities at the boundary.
We show in this paper that such impurities lead to dissipation in QH edges that is strongly enhanced when the impurity is tuned to resonance.
This dissipation can be observed in thermal imaging experiment, as has been recently reported \cite{Zeldov2019}.
Specifically, our theory explains the fascinating features observed in the experiment: the ring-shape structure of the thermal profile as a function of the tip position where a local potential is applied (Fig.\,\ref{fig:rings}).
Remarkably, this thermal profile is not accompanied with any electrical resistance (additional voltage drop). This should be contrasted to the case of a homogeneous edge (where nonlocal dissipation along the edge is associated with the voltage drop at the contacts), as well as to the resonant "supercollisions" in 2D materials \cite{Halbertal2017,TikhononvPRB18,KongPRB18} that lead to both local dissipation and local resistance. 

The manuscript is organized as follows.
We begin with the formulation of the model in Sec.\,\ref{sec:system}.
After that, we calculate the phonon emission rate induced by impurities along a QH edge state in Sec.\,\ref{sec:thermal_rings}. There we demonstrate that forward scattering at resonant impurities producse the experimentally observed thermal rings.
Based on the theory, we further address other experimental observations of Ref.\,\cite{Zeldov2019} in Sec.\,\ref{sec:experiment}.
In this section, we discuss the role of the edge reconstruction and propose a two-tip measurement to verify our theory.
We finally summarize our results and put forward further proposals in Sec.\,\ref{sec:summary}.
Technical details of calculations are relegated to appendixes.

\section{The System}
\label{sec:system}

We consider a sample which consists of a graphene layer encapsulated between two hexagonal Boron Nitride (h-BN) substrates \cite{Zeldov2019}. The system is placed under a perpendicular quantizing magnetic field that confines extended electron states to a strip with the width of the order of the magnetic length $l_m = \sqrt{1/eB}$ ($\hbar = 1$ for brevity) at the edge of the sample.
A metallic gate beneath the sample applies the back-gate voltage $V_{\text{BG}}$ to globally control the charge density. A superconducting tip (``SQUID-on-tip'' \cite{vasyukov,halbertal2016nanoscale,Halbertal2017}) is placed above the top h-BN layer. The role of the tip is twofold:
on one hand, it applies a tip voltage $V_{\text{tip}}$ which locally controls impurity levels; 
on the other hand, it measures the local temperature reflecting the energy dissipation rate \cite{halbertal2016nanoscale,Halbertal2017}.

We model the system by an effective single-particle Hamiltonian for electrons, 
which consists of four parts:
\begin{equation}
H = H_0 + H_{\text{T}} + H_{\text{dot}}
+ H_{\text{e-ph}}.
\label{eq:system_full_hamiltonian}
\end{equation}
Here, the first term $H_0$
describes the field-dependent motion of graphene electrons under the back-gate voltage.
We assume a sufficiently strong magnetic field such that the sample is in the integer QH regime where the QH edge electrons can be considered as 1D free particles in chiral channels \cite{GiamarchiBook,IhnBook}.
For simplicity, we focus on the case with filling factor $\nu = 2$, where the system contains only one topological chiral channel (after ignoring the spin degree of freedom), with the free Hamiltonian along the $x$ direction
\begin{equation}
H_0 = \sum_k \epsilon_k c_k^{\dagger} c_k,
\label{eq:1d_free_ham}
\end{equation}
where $\epsilon_k$ is the energy of the electron with momentum $k$, and $c_k^{\dagger}$ is its corresponding creation operator.

Strictly speaking, edge reconstruction in the graphene sample may lead to the emergence of additional (non-topological) counter-propagating channels \cite{SilvestrovEfetovPRB08}.
This can manifest itself in the local resistance features, as has been observed experimentally \cite{Zeldov2019}.
On the other hand, edge reconstruction does not essentially affect the mechanism of impurity-induced dissipation without resistance that is explored in the present paper.
Furthermore, the additional non-topological channels can be locally removed through the application of a local plunger gate that depletes quasiparticles or by the peculiar geometric confinement \cite{Zeldov2019}.
Experimentally these thermal rings are observed in regimes both with and without edge reconstruction \cite{Zeldov2019}.
We thus discard the edge reconstruction in the major part of the paper.

Realistic graphene samples contain resonant impurities that originate from irregularities and the missing dangling sites next to the material boundary.
These resonant impurities can be viewed as ``quantum dots'' that host electrons.
For simplicity, we consider a single impurity that couples to the chiral edge state
\begin{equation}
H_{\text{dot}} + H_{\text{T}} = \epsilon_d d^{\dagger} d + t \sum_{k}(c^{\dagger}_k d + h.c.),
\label{eq:impurity_and_coupling}
\end{equation}
where $H_{\text{T}}$ is the impurity-QH hybridization and $t$ is the momentum-independent coupling strength.
We assume that the energy difference between neighbouring impurity states is large enough such that only one impurity level is relevant, with the energy $\epsilon_d$.
For a fixed $V_{\text{tip}}$, the energy $\epsilon_d$ depends on the tip-impurity distance $l_{\text{ti}}$.
The first three terms of the Hamiltonian Eq.\,(\ref{eq:system_full_hamiltonian}) correspond to that of a resonant level model where an impurity is side-attached to a chiral edge state.

The fourth term of the Hamiltonian $H_{\text{e-ph}}$ is the electron-phonon interaction
\begin{equation}
\begin{aligned}
H_{\text{e-ph}} = & \iint d^2\vec{q} \sum_{k_1} \sum_{k_2} \frac{g_0 F(\vec{q})}{2\pi} \sqrt{\omega_q} M(k_1,k_2,q_x)\\
& \times  ( b_{\vec{q}} + b_{-\vec{q}}^{\dagger} ) \psi_{k_1}^{\dagger} \psi_{k_2},
\end{aligned}
\label{eq:eph_discrete}
\end{equation}
where
\begin{equation}
M(k_1,k_2,q_x) = \int dx \Psi_{k_1}^* (x) e^{i q_x x} \Psi_{k_2}(x)
\label{eq:m_matrix}
\end{equation}
is the phonon-induced scattering matrix element, and
$\psi^{\dagger}_k$ is the creation operator of the eigenstate with the corresponding wave function $\Psi_k(x)$, where $x$ is the coordinate along the edge.
For an impurity-free QH channel, $\psi^{\dagger}_k = c^{\dagger}_k$ and $\Psi_k(x) = e^{ikx}/\sqrt{L} \equiv \Psi_k^{(0)}(x)$, where $L$ is the size of the graphene sample.
With an impurity present, $\psi^{\dagger}_k$ instead creates a scattering state shown in Eq.\,(\ref{eq:scattering_operator}) below.
Because of the lack of backscattering along a QH edge, the scattering-state wave function simply acquires a phase shift after the scattering at the impurity.
In Eq.\,(\ref{eq:eph_discrete}), $g_0$ is the electron-phonon interaction strength, $\omega_q=s q$ the acoustic phonon dispersion with the sound velocity $s$, and $F(\vec{q})$ is the form factor \cite{SlizovskiyFalkoPRB17}
\begin{equation}
F(\vec{q}) = \int_{0}^{\infty} dy \sqrt{2} \sin(q_y y) \varphi^*_{k + q_x}(y) \varphi_{k }(y),
\label{eq:form_factor}
\end{equation}
with $\varphi_{k}(y)$ the $y$-direction electron wave-function
that is confined near the physical boundary within the range of $l_m$.
Such spatial confinement further leads to the relaxation of the momentum conservation along the $y$ direction \cite{SlizovskiyFalkoPRB17}.

Before proceeding with calculations, we discuss the features of the form factor defined in Eq.\,(\ref{eq:form_factor}).
First, the form factor strictly speaking depends also on the electron momentum $k$.
We neglect this dependence since $k \sim q_T s/v \ll q_T$, where $v \gg s$ is the electron Fermi velocity, and $q_T \equiv k_B T_{\text{el}}/s$ is the thermal momentum defined in terms of the electron temperature  $T_{\text{el}}$.
Second, because of the $\sin(q_y y)$ oscillation, the expression of the form-factor depends on the ratio between $|\vec{q}|$ and the inverse magnetic length $l_m^{-1}$.
More specifically, if $|\vec{q}| \gg l_m^{-1}$, $\sin(q_y y)$ of Eq.\,(\ref{eq:form_factor}) oscillates strongly so that $F(\vec{q}) = \chi_1/(q_y l_m)$, where $\chi_1$ is a $q_y$ independent prefactor.
In the opposite limit, $|\vec{q}| \ll l_m^{-1}$, $q_y y \ll 1$ so that $F(\vec{q}) = \chi_2 (q_y l_m)$.
Since the characteristic value of $|\vec{q}|$ is $q_T$, the character of electron-phonon scattering depends on the value of $q_T l_m$, which translates to the appearance of a $B$-dependent effective Bloch-Gr\"uneisen temperature $T_{\text{BG}} = s/(k_B l_m)$.
Below we study the dissipation in both limiting cases $q_T \gg l_m^{-1}$ and $q_T \ll l_m^{-1}$, i.e., the temperature above and below $T_{\text{BG}}$.

\section{Resonant Scattering and Thermal Rings}
\label{sec:thermal_rings}

With the Hamiltonian introduced above, we can calculate the energy dissipation induced by the presence of a resonant impurity.
In contrast to the 2D case, where energy dissipation originates from the particle scattering in all directions \cite{KongPRB18,TikhononvPRB18}, only forward scattering at the impurity is operative in the chiral QH edge.
This only introduces a phase factor in the wave function.

The scattering state in the presence of a single impurity thus has the general form
\begin{equation}
\Psi_k(x) = \frac{1}{\sqrt{L}} e^{ikx + i \Theta(x-x_0) \theta_k},
\label{eq:scattering_state_wf}
\end{equation}
where $\Theta(x-x_0)$ is the step function, $\theta_k$ is the momentum-dependent phase shift, and $x_0$ is the impurity position.
With the wave function (\ref{eq:scattering_state_wf}), the phonon-induced matrix element (\ref{eq:m_matrix}) becomes
\begin{equation}
\begin{aligned}
& M(k_1,k_2,q_x) = M_0(k_1,k_2,q_x) + M_{s}(k_1,k_2,q_x)\\
&= \delta_{k_1, q_x + k_2}
 -  i \frac{1}{L(k_1 - q_x + k_2) } [e^{-i(\theta_{k_1} - \theta_{k_2})} - 1],
\end{aligned}
\label{eq:effect_of_phase}
\end{equation}
where $M_0(k_1,k_2,q_x) $ is the impurity-free part. The effect of $M_0$ has already been studied in Ref.\, \cite{SlizovskiyFalkoPRB17}; it gives the following contribution to the energy dissipation rate:
\begin{equation}
\begin{aligned}
P_0 =
 \left\{\begin{array}{l} 
 \displaystyle
  \frac{g_0^2}{12 } \frac{s}{v^2} \frac{\chi_1^2}{l_m^2} k_B^2( T^2_{\text{el}} - T^2_{\text{lattice}} ),\ \ \ \ \ \ \ \ \ \ \ q_T\gg l_m^{-1}, 
  \\ 
  \\  
  \displaystyle
  \frac{4\pi^4 g_0^2}{63} \frac{s}{v^2} \frac{\chi_2^2 l_m^2}{ s^4 } k_B^6 ( T^6_{\text{el}} - T^6_{\text{lattice}} ),\ \ q_T\ll l_m^{-1}, \end{array}\right.
\end{aligned}
\label{eq:phonon_emission_free}
\end{equation}
where $T_{\text{lattice}}$ is the lattice temperature.
In Eq.\,(\ref{eq:phonon_emission_free}), $P_0$ has the dimension $\text{energy}/(\text{time} \times \text{length})$, which describes the amount of energy transferred between electrons and phonons per unit time and unit length.
The total dissipated power in the sample of size $L$ is $P_0 L$.

Below we investigate the effect of $M_s$ in two contrast scenarios, where electrons are scattered by either a scalar potential, or a resonant impurity.

\subsection{Energy dissipation for a scalar potential}
\label{sec:scalar_contribution}

We start by investigating the energy dissipation rate induced by a scalar potential.
Generally, a scalar potential locally changes the potential energy of quasiparticles. In the chiral 1D case, the effective Hamiltonian with a single scalar potential at $ x_0$ is simply
\begin{equation}
\begin{aligned}
H_{\text{eff}} & = H_0 + H_{\text{potential}}  = -i v \partial_x + V_s \delta (x - x_0),
\end{aligned}
\label{eq:scatterer_hamiltonian}
\end{equation}
where $V_s$ is the strength of the scalar potential and $v$ is the free electron Fermi velocity.

The straightforward solution of the Schrodinger equation with the Hamiltonian Eq.\,(\ref{eq:scatterer_hamiltonian}) gives us the box-normalized wave function
\begin{equation}
\Psi_k (x)  = e^{-i \frac{V_s}{v} \Theta(x-x_0)} \frac{1}{\sqrt{L}} e^{ikx},
\label{eq:wf_scatterer}
\end{equation}
where a momentum-independent phase shift $\theta_k = V_s/v$ occurs upon the scattering. Following Eq.\,(\ref{eq:effect_of_phase}), $M_s = 0$ and the scattering matrix will not be modified by the presence of the scalar potential. The scalar potential is thus trivial in the energy dissipation along a QH edge state.

The vanishing $M_s$ upon a scalar potential is a unique feature for 1D geometry, which should be contrasted to scattering at a scalar potential in 2D systems that creates a non-trivial energy dissipation rate \cite{SongLevitovPRL12,TikhononvPRB18}.
The irrelevance of a scalar potential to the creation of dissipation is in line with the intuitive expectation that the QH edge states are topologically protected from dissipation at local potentials.
However, when the scalar potential is replaced by a resonant impurity, this expectation breaks down, as we show below.

\subsection{Resonant impurity scattering}
\label{sec:resonant scattering}

The scattering of QH particles at a resonant impurity is in sharp contrast to that in the case a scalar potential in two major manners.

One one hand,
with a resonant impurity, electrons may be locally trapped in the ``quantum dot''. The wave function of the trapped electron creates an additional term $M_{\text{imp}}$ on top of Eq.\,(\ref{eq:effect_of_phase}). However, following the discussion of Appendix\,\ref{sec:imp_contribution}, $M_{\text{imp}}$ does not contribute to any dissipation when the point-like impurity couples to a single point of the QH edge state. We thus neglect $M_{\text{imp}}$ and focus on the effect of scattering states in the rest of the paper.

On the other hand, in comparison to the case with a scalar potential, the presence of a resonant impurity enables more scattering possibilities.
To see its effect, we derive the scattering state operator in the presence of a resonant impurity, following the technique introduced in Ref.~\cite{SchillerHershfieldPRB98}. The derivation details have been provided in Appendix\,\ref{sec:scatter_state_operator}, with the result
\begin{equation}
\!\Psi_k = \Psi_k^{(0)}+\frac{t}{\epsilon_k - \epsilon_d + i \Gamma} 
\sum_{k'} \frac{t\  \Psi_{k'}^{(0)}}{\epsilon_k - \epsilon_{k'} + i \eta},
\label{eq:scattering_operator}
\end{equation}
where $\Psi_k^{(0)}$ is the unperturbed electronic plane wave, $\Gamma$ 
is the level broadening \cite{Bruus-Flensberg}, and $\eta $ is a positive infinitesimal. 

In Eq.\,(\ref{eq:scattering_operator}), the first and second terms correspond to the wave function of the free and the scattered states, respectively. For an infinite chiral channel with spectrum $\epsilon_k=vk-\mu$ (with $\mu$ the chemical potential), the integral over $k'$ in wave-function (\ref{eq:scattering_operator}) is given by its residue $\epsilon_{k'}=\epsilon_k$. Then the scattering state wave function (\ref{eq:scattering_state_wf}) can be conveniently written in terms of the $k$-dependent phase shift of the plane wave at $x>x_0$:
\begin{equation}
\theta_k = -2  \arctan\frac{\Gamma}{\epsilon_k-\epsilon_d}.
\label{shift-k}
\end{equation} 
The main difference of this phase shift from the one in the case of a scalar potential, Eq.~(\ref{eq:wf_scatterer}), is its momentum dependence. It is this feature of the phase shift 
(\ref{shift-k}) that yields dissipation without resistance in the case of resonant impurity scattering.

\subsection{Dissipation induced by resonant impurities}
\label{sec:supercollisions}

With the scattering phase shift (\ref{shift-k}), the matrix element $M_s$ produced by a resonant impurity becomes (see Appendix \ref{sec:scatt_state_dissipation}):
\begin{equation}
M_s(k_1, k_2,q_x) 
 \approx \frac{2 \Gamma}{- \epsilon_d - i\Gamma } \cdot \frac{e^{i q_x L/2} - e^{i q_x x_0}}{ L  ( q_x - \omega_q/v)},
\label{eq:m0}
\end{equation}
Here we have used the energy conservation prescribing that $v(k_2 - k_1) = \omega_q$ for the process of phonon emission and have taken the limit $\epsilon_d,\Gamma \ll s q_T$.
With Eq.\,(\ref{eq:m0}), we calculate the energy dissipation rate at $x > x_0$ in two limiting cases:
\begin{equation}
\begin{aligned}
\! P_{\text{imp}} \!
& = \!\left\{\begin{array}{l}\! \displaystyle \!\frac{g_0^2}{3}\! \frac{s}{v^2}\! \frac{\chi_1^2}{l_m^2} k_B^2 ( T^2_{\text{el}} - T^2_{\text{lattice}} ) \frac{\Gamma^2}{\Gamma^2 + \epsilon_d^2},  
\\ 
\\ \! \displaystyle \!\frac{16 \pi^4 g_0^2}{63}\! \frac{s}{v^2}\! \frac{\chi_2^2 l_m^2}{ s^4 } k_B^6( T^6_{\text{el}} - T^6_{\text{lattice}} ) \frac{\Gamma^2}{\Gamma^2 + \epsilon_d^2},
\end{array}\right. 
\end{aligned}
\label{eq:diss_rate_ls}
\end{equation}
for $q_T\gg l_m^{-1}$ and $q_T\ll l_m^{-1}$, respectively.
The impurity-induced energy dissipation rate Eq.\,(\ref{eq:diss_rate_ls}) is the central result of this paper.
It shows a resonant feature:
the energy dissipation rate reaches its maximum when $\epsilon_d = 0$, and decreases in a Lorentzian manner with the half-width $\Gamma$.
This resonant feature leads to the experimentally observed thermal rings. Specifically, at a certain tip-impurity distance, the impurity level is fine-tuned by the tip voltage to resonance, which gives rise to the enhanced dissipation according to Eq.\,(\ref{eq:diss_rate_ls}).

The resonant character of dissipation in Eq.\,(\ref{eq:diss_rate_ls}) bears similarity with resonant supercollisions in 2D graphene bulk \cite{KongPRB18,TikhononvPRB18}.
However, the chiral 1D case is distinct in several respects.
These peculiarities are related to the chiral 1D nature of the edge states.

First, the 2D theory implies the increase of resistance associated with the increase
of dissipation. In contrast to that, dissipation introduced by resonant scatterers in 1D edges is not accompanied by any local voltage drop (i.e., any local change of the resistance).
This leads to an important conclusion: in a chiral QH channel, forward scattering at a resonant impurity gives rise to a finite energy dissipation rate, despite of the topological protection of the Hall conductance.

Second, the dissipation rate (\ref{eq:diss_rate_ls}) does not depend on $x$ for $x > x_0$, which means that the impurity-induced dissipation in 1D is ``global''. This result is also distinct from that in 2D, where the major energy dissipation occurs locally near the scattering impurity \cite{SongLevitovPRL12,KongPRB18,TikhononvPRB18}.
Indeed, in 1D edges, scattering states do not decay with respect to $x$, in contrast to the 2D geometry. 
As a result, the dissipation induced by a single impurity in 1D edge becomes extensive:
the entire QH edge past the impurity participates in the energy dissipation, leading to its ``global'' feature.

\begin{figure}[ht]
\centering
\includegraphics[width = 0.5 \textwidth]{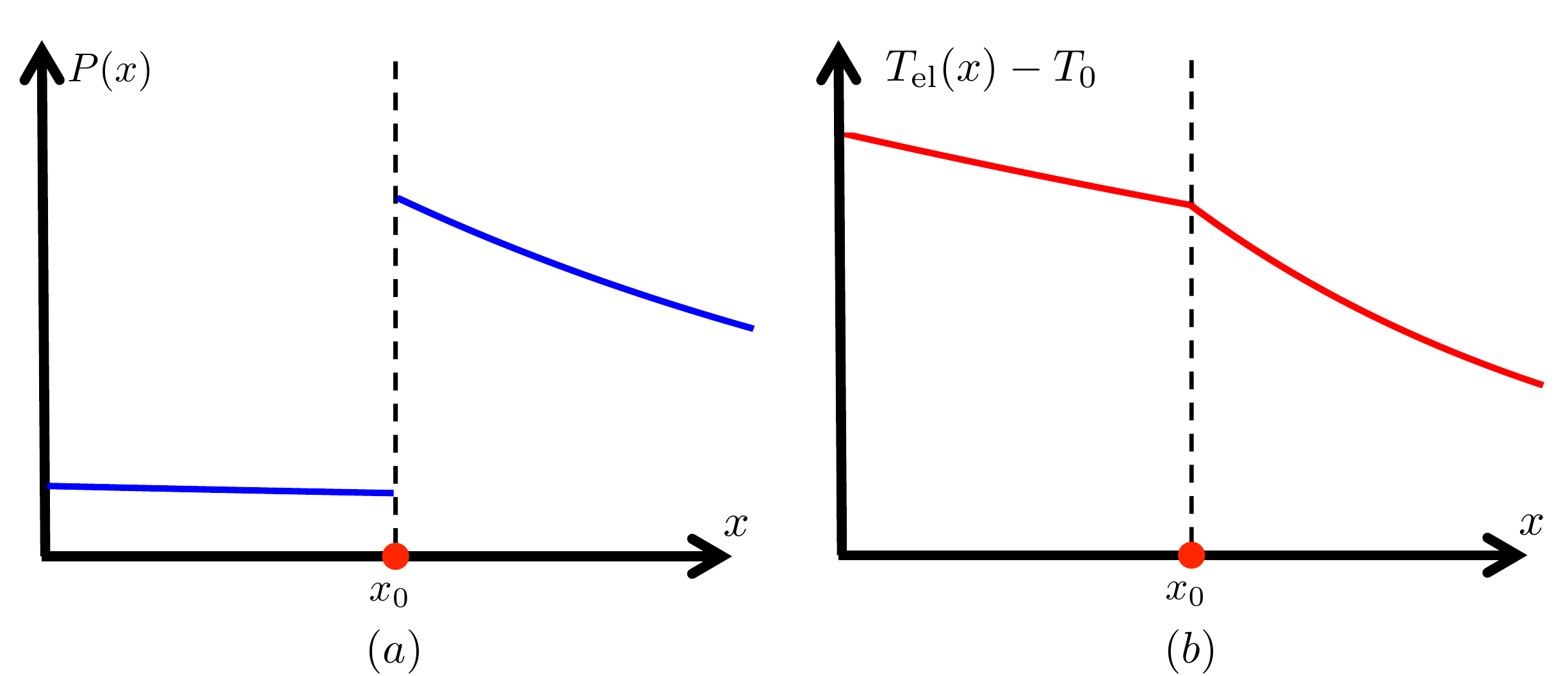}
\caption{The spatial dependence of (a) the energy dissipation rate, and (b) the electron temperature. An on-resonance impurity is placed at $x = x_0$. The temperature $T_0$ is the background temperature which is assumed to be constant. We make the plots following the high-temperature expression of Eq.\,(\ref{eq:diss_rate_ls}).}
\label{fig:dissipation_rate}
\end{figure}

Strictly speaking, the electron temperature $T_{\text{el}}$ decreases after the phonon emission.
The strong temperature dependence of Eq.\,(\ref{eq:diss_rate_ls}) then indicates an accompanied decrease in the phonon emission rate.
With the standard technique (see Ref.\,\cite{SlizovskiyFalkoPRB17}, for instance) and the steady state assumption, the temperature gradient becomes
\begin{equation}
\frac{\partial k_B T_{\text{el}}(x)}{\partial x} = - \frac{P(x)}{C v},
\label{eq:temperature_variation}
\end{equation}
where $C = \pi k_B T_{\text{el}}/(6\hbar v)$ is the specific heat of the QH edge state.
The dissipation rate $P(x) = P_0(x)$ when $x<x_0$ and $P(x) = P_0(x) + P_{\text{imp}}(x)$ otherwise.
Based on Eq.\,(\ref{eq:temperature_variation}), we plot the energy dissipation rate and the electron temperature as a function of the position in Fig.\,\ref{fig:dissipation_rate}, where $T_0$ is the background temperature, and $x_0$ is the impurity position.
The decreasing electron temperature ensures that the total dissipated power does not diverge with the system size.

\section{Linking to the experiment}
\label{sec:experiment}

With the impurity-induced dissipation rate Eq.\,(\ref{eq:diss_rate_ls}), here we discuss a connection between our theory and the experimental data.

\subsection{Estimates for thermal rings}

As has been pointed out above, the resonant feature of Eq.\,(\ref{eq:diss_rate_ls}) leads to the experimentally observed thermal rings, with their centers located at the resonant impurity sites.
Since the half-width of a dissipation peak is $\sim \Gamma$, these rings have the thickness $\sim r_{\text{ring}} \Gamma /V_{\text{tip}}$, which is much smaller than the ring radius $r_{\text{ring}} $.
This explains the sharp contrast of the experimentally observed thermal rings \cite{Zeldov2019}.

On the quantitative level, we evaluate the temperature enhancement induced by the impurity. To begin with,
the experiment is carried out with the background temperature $T_{0} \approx 4.2 K$ and under the magnetic field $B \approx 1 T$.
Based on Ref.\,\cite{SlizovskiyFalkoPRB17}, we focus on the LA phonons with the longitudinal sound velocity $s = 2.2 \times 10^4 m/s$. We also take the interaction parameter for LA phonons $g_0 \approx 1.7 \times 10^{-19} J\cdot s /\sqrt{kg} $. Finally, we take the form factor $\chi \approx 0.96$ and the graphene Fermi velocity $v = 10^6 m/s$. 
The derivation of the impurity induced temperature enhancement requires a solution of the heat diffusion equation
\begin{equation}
- \kappa_{\text{lattice}} \bigtriangledown^2 T_{\text{lattice}}  = P(x) \delta(y) - \gamma_0 (T_{\text{lattice}} - T_0),
\label{eq:diffusive}
\end{equation}
where $\gamma_0$ quantifies the coupling to the bath and $\kappa_{\text{lattice}}$ is the phonon conductivity.
In Eq.\,(\ref{eq:diffusive}), we simply the model with the assumption that only electrons right at the boundary emit phonons.
In reality, the phonon emission involves electrons in the edge state with the width $\sim l_m$.
Since phonons propagate in 2D, we evaluate the value of $\kappa_{\text{lattice}}$ following the 2D equations \cite{KongPRB18},
\begin{equation}
\kappa_{\text{lattice}} = \frac{9 Z\zeta(3) k_B^2 T_0}{\hbar} \approx 2 \times 10^{-8} \ W/K,
\label{eq:kappa}
\end{equation}
where $Z \approx 200$ is the number of atom layers of the graphene-hBN system with the width $\approx 60 nm$ \cite{Zeldov2019}.

Experimentally, electron temperature enhances near the constriction due to the energy input.
Following the heat transport equation, and ignore the phonon emission (since it is negligible in comparison to the rate of the input energy), we arrive at the expression of the electron temperature at the constriction $T_{\text{el}} (0)$
\begin{equation}
[T_{\text{el}}^2(0) - T_0^2 ] = \frac{12}{N} \frac{\hbar \eta W_{\text{input}}}{\pi k_B^2},
\label{eq:tel0}
\end{equation}
where $N = 2$ for two graphene edges, and $W_{\text{input}} = 10 nW$ is the energy input. We take the efficiency parameter $\eta = 0.5$, assuming that half of the input energy is transferred into the electron temperature at the source (the other half is dissipated at the drain).
With Eq.~\eqref{eq:tel0}, we get the electron temperature at the constriction $T_{\text{el}} (0) \approx 50 \ K$.
Based on Ref.\,\cite{Zeldov2019}, without the contribution from resonant impurities, the measured edge temperature is 
\begin{equation}
\delta T_{\text{edge}} = T_{\text{lattice}} - T_0 \approx 150 \mu K
\end{equation}
higher than the background temperature.
We thus get 
\begin{equation}
\gamma_0 = P_0/(\delta T_{\text{edge}} l_m) \approx 5 \times 10^5 W/(m^2 \cdot K),
\end{equation}
which is reasonably close to the experimental result \cite{ChenAPL09}.

With these values of parameters, we begin to calculate the phonon temperature following Eq.\,(\ref{eq:diffusive}).
Strictly speaking, $P(x) \propto T_{\text{el}}^2(x) - T_{\text{lattice}}^2 (x)$ and the spatial dependence of the electron temperature should be considered.
However, in the presence of an on-resonance impurity, electron temperature cools down [following Eq.~\eqref{eq:temperature_variation}] with the characteristic cooling length \cite{SlizovskiyFalkoPRB17}
\begin{equation}
l_{\text{cool}} = \frac{\pi v^2 l_m^2}{2\hbar g_0^2 s \chi_1^2} \approx 20\  \mu m,
\label{eq:cooling length}
\end{equation}
which is of the same order of the system size.
We thus ignore the spatial dependence of electron temperature $T_{\text{el}}(x) \approx T_{\text{el}}(0)$ in the evaluation that follows.

Meanwhile, since $T_{\text{el}}^2 (0) \gg T_{\text{lattice}}^2 (x) \approx T_0^2$, we approximately treat both $P_0$ and $P_{\text{imp}}$ as constant.
We further define dimensionless parameters $\tilde{ T} \equiv[T_{\text{lattice}} - T_0] /\delta T_{\text{edge}}$, $X \equiv x/l_d$, and $Y \equiv y/l_d$, where 
\begin{equation}
l_d \equiv\sqrt{\kappa_{\text{lattice}/\gamma_0}}\approx 200 n m
\end{equation}
is the typical traveling distance of phonons before they enter the bath.
With these dimensionless parameters, we rewrite Eq.~\eqref{eq:diffusive} as
\begin{equation}
\begin{aligned}
 - (\partial_{X}^2 + \partial_{Y}^2) \tilde{T}
& = \left\{\begin{array}{l} \displaystyle \delta (Y) -  \tilde{T} ,  
\ \ \ \ \ \ \ \  \ \ \ X < 0, 
\\ 
\\  \displaystyle 5 \delta (Y) -  \tilde{T},
\ \ \ \ \ \ \ \ \ X > 0.
\end{array}\right. 
\end{aligned}
\label{eq:normalized_equation}
\end{equation}

\begin{figure}[ht]
\centering
\includegraphics[width = 0.5 \textwidth]{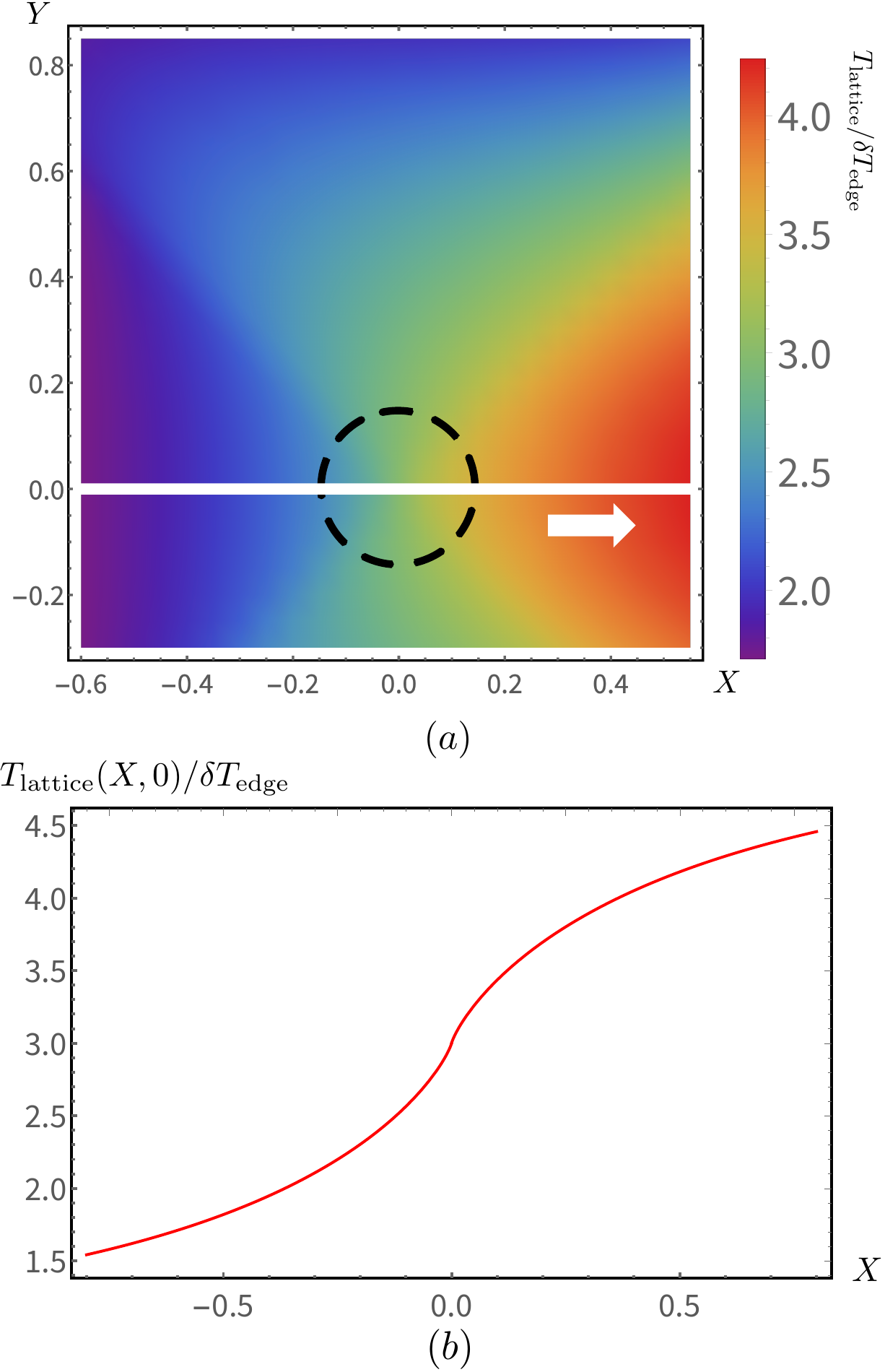}
\caption{The temperature profile $T_{\text{lattice}} (X,Y)$ obtained by solving Eq.~\eqref{eq:normalized_equation}. (a) The 2D temperature profile through numerical integral. The black dashed line refers to the tip position where a thermal ring is detected. The edge state (represented by the white solid line) is propagating rightward (the white arrow). (b) The plot of the analytical result Eq.~\eqref{eq:temperature_profile_x} when $Y = 0$.}
\label{fig:temperature_profile}
\end{figure}

It is conventional to solve Eq.~\eqref{eq:normalized_equation} by dividing the edge into point-like sources, with the solution
\begin{equation}
T_{\text{lattice}}(x,y) - T_0 = \int_{-\infty}^{\infty} dx' S(x') \frac{K_0(r)}{\pi}
\label{eq:temperature_profile}
\end{equation}
where $K_n(r)$ is the modified Bessel function of the second kind, with $r = \sqrt{(x-x')^2 + y^2}$ the distance to a point-like source. The function $S(x) = 1$ for $x<0$ and $S(x) = 5$ otherwise.
At the boundary $Y=0$, we have the analytical expression of Eq.~\eqref{eq:temperature_profile} 
\begin{widetext}
\begin{equation}
\begin{aligned}
\tilde{T}(X,0)
& = \left\{\begin{array}{l} \displaystyle 1 + 2 [1 + X K_0(-X) L_{-1}(X) - X K_1(-X) L_{0}(X)],  
\ \ \ \ \ \ \ \  \ \ \ X < 0, 
\\ 
\\  \displaystyle 3 + 2 X [K_0(X) L_{-1}(X) + K_1(X) L_0(X)],
\ \ \ \ \ \ \ \  \ \ \ \ \ \ \ \ \ \ \  \ \ \ \ \ \ \ \ \  X > 0,
\end{array}\right. 
\end{aligned}
\label{eq:temperature_profile_x}
\end{equation}
\end{widetext}
where $L_n(x)$ is the modified Struve function.
For $Y\neq 0$, we integrate Eq.~\eqref{eq:temperature_profile} numerically, with the result presented in Fig.\,\ref{fig:temperature_profile}\textcolor{red}{a}.
Notice that the non-locality of the phonon emission introduces the anisotropy of a thermal ring.
However, the complete temperature profile is experimentally inaccessible with the single-tip measurement.

Experimentally, thermal rings have typical radius around $30 \sim 100 nm$. For a thermal ring with radius $50 nm$, the impurity induced phonon temperature enhancement is around $250 \mu K$ to $350 \mu K$, depending on the tip position.
This value agrees quite well with the experimental data.

Before the end of the subsection, we emphasize that we obtain the results Eq.~\eqref{eq:temperature_profile} and Fig.\,\ref{fig:temperature_profile} after neglecting the electron temperature variation.
On a larger scale, after arriving at its peak value around $x = l_d$, the phonon temperature $T_{\text{lattice}}$ decreases simultaneously with the decreasing $T_{\text{el}}$.
This decreasing in $T_{\text{el}}$ becomes manifest when $x \gg l_{\text{cool}}$, where the phonon temperature enhancement $T_{\text{lattice}} - T_0$ also (almost) vanishes.

\subsection{Role of the edge reconstruction}

The major result Eq.~\eqref{eq:diss_rate_ls} only requires the existence of forward scattering at resonant impurities.
However, edge reconstruction should be included to fully describe all the findings of Ref.\,\cite{Zeldov2019}. 

In the experiment \cite{Zeldov2019}, bias is applied to the source while the drain is grounded.
In this situation, our analysis with a full chiral edge yields thermal rings only on the edge segment downstream from the source.
This agrees with the experimental observation for high filling factors.
At the same time, for the case of low filling factors (close to unity), experiment reveals dissipation rings also upstream from the source.
The possible explanation of this is related to the edge reconstruction, as also proposed in Ref.\,\cite{Zeldov2019}.
Indeed, if the equilibration length in the reconstructed edge is comparable to the system size, energy can also propagate from the source upstream over the corresponding segment of the edge.
This will also heat the outmost channel on the segment. The scattering of this channel at the impurities near the sample boundary will produce the rings observed.

Meanwhile, backscattering between counter-propagating channels in the edge-reconstructed area leads to extra amount of dissipation.
In strong contrast to the resonance-induced thermal rings discussed in this paper, the backscattering-induced thermal signal (i) does not generically display the ring shape, and (ii) is strongly correlated with the scanning gate signal which measures the longitudinal Hall resistance.
Their strong correlation will be explained elsewhere.

Finally, in the edge reconstructed area, the backscattering-induced dissipation strongly depends on the equilibration between counter-propagating channels.
As a possible extension of the experiment, we thus propose that the scanning-tip technique can be employed to systematically explore the equilibration feature of other systems that consist of counter-propagating chiral channels.
Fractional QH systems, where a direct measurement of the equilibration pattern is missing, are possible candidates.

\subsection{Non-local dissipation in experiment}

As the central prediction of this paper, the non-locality of the impurity-induced dissipation cannot be directly measured by a single tip in the current experiment of Ref.\,\cite{Zeldov2019}. We thus propose a two-tip experiment to investigate the global feature of the dissipation in 1D. In this experiment, one tip (the detector) measures the sample temperature and the other (the tuning tip) tunes the impurity on-resonance.

Following the mechanism that produces dissipation in 2D \cite{TikhononvPRB18}, the scattering wave function decays in the $\sim r^{-2}$ manner where $r$ is the distance from the resonant impurity.
The resulting temperature enhancement induced by supercollisions involving this impurity displays a 
fast-decaying profile as a function of the distance from the impurity position. This profile can be probed by the detector in the two-tip measurement.

In contrast, in the QH regime, if the tuning tip tunes one boundary impurity on-resonance, the edge temperature measured by the detector follows the pattern shown in Eq.\,(\ref{eq:temperature_profile}), as has been plotted in Fig.\,\ref{fig:temperature_profile}.
More specifically, when the detector moves from the upstream side to the downstream side of the on-resonance impurity, it detects an abrupt temperature enhancement; otherwise the temperature measured by the detector varies slowly and smoothly.
As a comparison, the single-tip measurement only detects temperature of points on the black dashed ring, and is thus unable to detect the full temperature profile.
The two-tip measurement is thus capable to verify our theory.

Finally, the two-tip measurements can also verify our explanation on the upstream thermal rings. Experimentally, the tuning tip should be placed at the position where one upstream thermal ring is produced in Ref.\,\cite{Zeldov2019}. At such position, one impurity at the upstream side of the source has been tuned on-resonance by the tuning tip. The detector should then detect an extensive heat production along the downstream direction of the on-resonance impurity. Otherwise, heat production should be equally detected on both sides of the impurity.

\section{Summary and outlook}
\label{sec:summary}

We have studied dissipation in a chiral edge of a QH sample in the presence of a resonant impurity. 
In strong contrast to an expectation that forward scattering is irrelevant to energy dissipation, we have shown that the forward scattering within a single chiral QH channel at a resonant impurity produces a non-trivial dissipation enhancement.
We have found that the phonon emission rate is maximized through supercollisions at an on-resonance quantum dot.
This enhancement, which is related to the momentum-dependent phase shift, leads to a finite impurity-induced temperature peak in local thermal nano-imaging, thus explaining the appearance of thermal (``entropy'') rings observed in a very recent experiment \cite{Zeldov2019}.

These thermal rings are distinct from those in 2D systems in the following respects: (i) the dissipation in a 1D chiral edge is global because the wave function modification induced by the impurity remains along the entire edge and (ii) the dissipation pattern is uncorrelated with any local resistance variations.

Before closing the paper, we discuss several prospective directions for future research related to our work.
As mentioned above, the experiment has observed the strong dependence of equilibration length on the filling factor. it is thus interesting to include systematically the edge reconstruction and inter-channel equilibration into the theoretical study of dissipation in QH graphene samples.

As a further direction, it will also be interesting to study the impurity-induced dissipation in other topological systems with complex edge structures, including fractional QH and spin QH systems.
These systems may naturally provide counter-propagating edge modes and are characterized by strong correlations, so that the impurity-induced dissipation feature may be distinct from that of the integer QH system studied in this paper.

Finally, in this paper, we assumed that only one impurity is tuned on-resonance by the tip.
However, one can imagine that several impurities can be tuned on-resonance simultaneously under certain circumstances. The exploration of dissipation in this case is another prospective topic.

\begin{acknowledgments}
We are grateful to A. Aharon-Steinberg, A. Marguerite, and E. Zeldov for insightful discussions and sharing the unpublished experimental data with us. We thank Y. Gefen, A. Dmitriev, V. Kachorovskii, K. Tikhonov, and C. Sp\aa{}nsl\"att for interesting discussions. The work was supported by the FLAGERA
JTC2017 Project GRANSPORT through the DFG grant No. GO 1405/5, by the DFG grant No. MI  658/10-1, and by 
the German-Israeli Foundation (GIF Research Grant No. I-1505-303.10/2019).
\end{acknowledgments}

\begin{appendix}

\section{Effect of a Resonant Impurity: the Scattering States Operators}
\label{sec:scatter_state_operator}

In this section we present the details in getting the scattering state wave function Eq.~\eqref{eq:scattering_operator} following the "$L$ operator" method \cite{SchillerHershfieldPRB98}. We begin with the Hamiltonian Eqs.\,(\ref{eq:1d_free_ham}) and (\ref{eq:impurity_and_coupling})
\begin{equation}
\begin{aligned}
H  & = \sum_k \epsilon_k c^{\dagger}_k c_k + t \sum_k ( c^{\dagger}_k d + d^{\dagger} c_k ) + \epsilon_d d^{\dagger} d \\
& = H_0 + H_{\text{T}} + H_{\text{dot}},
\end{aligned}
\label{eq:system_ham}
\end{equation}
where the electron-phonon interaction $H_{\text{e-ph}}$ has been ignored.
For later convenience, here we define operators $\hat{L}_n$
\begin{equation}
\hat{L}_n \hat{O} = [\hat{O}, H_n],
\label{eq:ln_ops}
\end{equation}
where $\hat{O}$ is any operator and $n \in \left\{ 0,\text{T},\text{dot} \right\}$ for the three parts of the Hamiltonian. We further define $\hat{L} = \hat{L}_0 + \hat{L}_{\text{T}} + \hat{L}_{\text{dot}}$.

The target is to rewrite Hamiltonian Eq.\,(\ref{eq:system_ham}) into the effective one $H'= \sum_k \epsilon_k \psi_k^{\dagger} \psi_k$ with the scattering state operators $\psi_k$ that are linear combinations of bare lead operators $c_k'$ and the impurity operator $d$. The scattering state operators satisfy the commutation relation \cite{SchillerHershfieldPRB98}
\begin{equation}
[ \psi^{\dagger}_k , H' ] = -\epsilon_k \psi^{\dagger}_k + i\eta (c^{\dagger}_{k} - \psi^{\dagger}_k),
\label{eq:scattering_state_commutator}
\end{equation}
where $\eta = 0^+$ is a positive infinitesimal. With definition of the $\hat{L}$ operator, Eq.\,(\ref{eq:scattering_state_commutator}) can be rewritten as
\begin{equation}
\psi^{\dagger}_k = c^{\dagger}_k + \frac{t}{\hat{L} + \epsilon_k + i\eta} d^{\dagger}.
\label{eq:commutator_rewritten}
\end{equation}
Apparently, we need to express the second term of Eq.\,(\ref{eq:commutator_rewritten}) in terms of bare operators. Straightforwardly, following the definition Eq.~\eqref{eq:ln_ops}, it becomes
\begin{equation}
\begin{aligned}
&\frac{1}{\hat{L} + \epsilon_k + i\eta} d^{\dagger}
 = \frac{1}{\epsilon_k + i\eta} d^{\dagger} +\\
& \frac{1}{\hat{L} + \epsilon_k + i\eta} d^{\dagger} \frac{\epsilon_d}{\epsilon_k + i\eta} 
- \frac{1}{\epsilon_k + i\eta} \frac{1}{\hat{L} + \epsilon_k + i\eta} \hat{L}_{\text{T}} d^{\dagger}.
\end{aligned}
\label{eq:ld_calculation}
\end{equation}
In the derivations above, we have used the fact that $\hat{L}_{\text{dot}} d^{\dagger} = - \epsilon_d d^{\dagger}$. The last term of Eq.\,(\ref{eq:ld_calculation}) can be calculated through
\begin{widetext}
\begin{equation}
\begin{aligned}
\frac{1}{\hat{L} + \epsilon_k + i\eta} \hat{L}_{\text{T}} d^{\dagger} & = \frac{1}{\hat{L}_0+\hat{L}_{\text{dot}} +\epsilon_k+i\eta} \hat{L}_{\text{T}} d^{\dagger} - \frac{1}{\hat{L} +\epsilon_k+i\eta} \hat{L}_{\text{T}} \frac{1}{\hat{L}_0+\hat{L}_{\text{dot}} +\epsilon_k+i\eta} \hat{L}_{\text{T}} d^{\dagger} \\
& = - \sum_{k'} \frac{t}{\epsilon_k - \epsilon_{k'} + i\eta} c^{\dagger}_{k'} + \frac{1}{\hat{L}+\epsilon_k+i\eta} \hat{L}_\text{T} \sum_{k'} \frac{t}{\epsilon_k - \epsilon_{k'} +i\eta} c^{\dagger}_{k'} 
\\
& = -\sum_{k'} \frac{t}{\epsilon_k - \epsilon_{k'} + i\eta} c^{\dagger}_{k'} + \frac{i \Gamma}{\hat{L} + \epsilon_k + i\eta} d^{\dagger} ,
\end{aligned}
\label{eq:ll1d}
\end{equation}
\end{widetext}
where we have used 
$$\hat{L}_{\text{T}} d^{\dagger} = - V\sum_k c^{\dagger}_k, 
\ \quad \hat{L}_{\text{dot}} c^{\dagger}_k = 0, \ \quad 
\hat{L}_0 c^{\dagger}_{k} = -\epsilon_k c^{\dagger}_{k}.$$
In the last step of Eq.\,(\ref{eq:ll1d}), we have used the continuous momentum 
$k$ and a large-band approximation:
\begin{equation}
\begin{aligned}
&\ \ \ \ \sum_{k'} \frac{t^2}{\epsilon_{k} - \epsilon_{k'} + i\eta}  = \rho \int d \epsilon_{k'} \frac{t^2}{\epsilon_{k} - \epsilon_{k'} + i\eta} \\
& = - \frac{\Gamma}{\pi} \ln \Bigg| \frac{D+\epsilon_k}{D - \epsilon_k} \Bigg| - i \Gamma \approx -i\Gamma.
\end{aligned}
\label{eq:large_band_limit}
\end{equation}
Here $\rho$ is the density of state $\rho = L/(2\pi v)$ and $\Gamma = \pi \rho t^2$ is the dot level broadening. At the large band limit, the real part of Eq.\,(\ref{eq:large_band_limit}) vanishes.

With Eqs.~\eqref{eq:commutator_rewritten}, (\ref{eq:ld_calculation}), and (\ref{eq:ll1d}) combined, we
finally get the scattering state operators
\begin{equation}
\psi^{\dagger}_k\! =\! c^{\dagger}_k \!+ \!\frac{t}{\epsilon_k - \epsilon_d \!+\! i \Gamma} 
\left( d^{\dagger} +  \sum_{k'}\! \frac{t}{\epsilon_k - \epsilon_{k'} + i\eta} c^{\dagger}_{k'} \right).
\label{eq:scattering_ops_expression}
\end{equation}
After ignoring the part of the wave function localized on the impurity (the term with $d^{\dagger}$),
we arrive at Eq.~\eqref{eq:scattering_operator}.

\section{Energy emission rate induced by a resonant impurity}
\label{sec:scatt_state_dissipation}

With the scattering state wave function derived in Sec.\,\ref{sec:scatter_state_operator}, we calculate the energy emission rate induced by a resonant impurity.
In this section only the contribution from $M_{s}$ will be calculated.
The contribution from $M_{\text{imp}}$ will be evaluated in Sec.\,\ref{sec:imp_contribution}.
Based on Eq.~\eqref{eq:scattering_state_wf}, the scattering at a resonant impurity located at $x=x_0$ 
induces a momentum dependent phase shift $\theta_k$.
The expression of $\theta_k$ has been provided by Eq.\,(\ref{eq:m0}).

The phonon-induced matrix element calculated with the wave function Eq.\,(\ref{eq:scattering_state_wf}) becomes
\begin{equation}
\begin{aligned}
&\frac{1}{L}  \int_{-\frac{L}{2}}^{\frac{L}{2}} dx e^{i(-k_2 + q_x + k_1 )x} e^{-i \Theta(x - x_0) (\theta_{k_2} - \theta_{k_1}) } 
\\
&= \delta_{k_2,k_1 + q_x}\!+\![ e^{-i(\theta_{k_2} - \theta_{k_1})}\! -\! 1 ]\! 
\int_{x_0}^{\frac{L}{2}}\! dx \frac{e^{i(-k_2 + q_x + k_1 )x}}{L},
\end{aligned}
\label{eq:e-ph_element}
\end{equation}
where the first term is the impurity-free contribution and the second term comes from the electron scattering at the impurity.

In the limit $\Gamma,\epsilon_d \ll k_B T_{\text{el}}$, we can approximate 
$$\exp[-i(\theta_{k_2} - \theta_{k_1})] - 1 \approx -2\Gamma/(\epsilon + i\Gamma).$$
To study the effect of an impurity on the energy dissipation processes, we need to evaluate the integral
\begin{equation}
\begin{aligned}
&\frac{2\Gamma}{-\epsilon - i\Gamma}\int_{x_0}^{L/2} dx \frac{1}{L} e^{i(-k_2 + q_x + k_1 )x} \\
 =& \frac{2\Gamma}{-\epsilon - i\Gamma} \frac{e^{i(-k_2 + q_x + k_1)\frac{L}{2}} - e^{i(-k_2 + q_x + k_1)x_0}}{iL(-k_2 + q_x + k_1)}\\
 \approx & \frac{2\Gamma}{-\epsilon - i\Gamma} \frac{e^{i\frac{q_x L}{2}} - e^{i q_x x_0}}{i q_x L },
\end{aligned}
\label{eq:function_overlap}
\end{equation}
where in the second line we have used the fact that typically $q_x \gg k_1, k_2$.
The absolute square of Eq.\,(\ref{eq:function_overlap}) reads:
\begin{equation}
\frac{4\Gamma^2}{\epsilon_d^2 + \Gamma^2}\frac{2 - 2\cos[q_x (L/2 - x_0) ]}{q_x^2L^2},
\label{eq:overlap_expression}
\end{equation}
With Eq.~(\ref{eq:overlap_expression}) we calculate the phonon absorption rate induced by the impurity:
\begin{widetext}
\begin{equation}
\begin{aligned}
\Gamma_a & = \iint d^2\vec{q} \sum_{k_1} \sum_{k_2} \frac{g_0^2}{(2\pi)^2} F^2(q_y) \omega_q \frac{2 - 2\cos[q_x (L/2 - x_0) ]}{q_x^2L^2} 
[ 1 - n_F(\epsilon_{k_1}) ] n_F(\epsilon_{k_2}) N_B^{\text{lattice}} (\omega_q) \frac{4\Gamma^2}{\epsilon_d^2 + \Gamma^2} \delta (\epsilon_{k_1} - \epsilon_{k_2} - \omega_q ) 
\\
& = \iint d^2\vec{q} \frac{g_0^2 L^2}{(2\pi)^4 v^2} F^2(q_y) \omega_q^2 N_B^{\text{lattice}} (\omega_q) [ N_B^{\text{el}} (\omega_q) + 1 ] \frac{2 - 2\cos[q_x (L/2 - x_0) ]}{q_x^2L^2}\frac{4\Gamma^2}{\epsilon_d^2 + \Gamma^2}.
\end{aligned}
\label{eq:impurity_absorption_rate}
\end{equation}
\end{widetext}
Here, $n_F$ and $N_B^{\text{lattice}}$ are the electron and phonon distribution functions, 
and $N_B^{\text{el}}$ describe the Bose distribution with temperature $T_{\text{el}}$. 

The energy dissipation rate in the high temperature limit $q_T \gg l_m^{-1}$  reads
\begin{equation}
\begin{aligned}
W_{\text{impurity}}& = 2 \Gamma (2) \zeta (2) \frac{g_0^2}{\pi^2} \frac{s}{v^2} \frac{\chi_1^2}{l_m^2} k_B^2[ T^2_{\text{el}} - T^2_{\text{lattice}} ] 
\\&  \times \frac{\Gamma^2}{\Gamma^2 + \epsilon_d^2}(L/2 - x_0).
\end{aligned}
\label{eq:expression_w1}
\end{equation}
Similarly, in the opposite limit, it becomes
\begin{equation}
\begin{aligned}
W_{\text{impurity}} & = 2 \Gamma (6) \zeta (6) \frac{g_0^2}{\pi^2} \frac{s}{v^2} \frac{\chi_2^2 l_m^2}{ s^4 } k_B^6[ T^6_{\text{el}} - T^6_{\text{lattice}} ] \\
&  \times \frac{\Gamma^2}{\Gamma^2 + \epsilon_d^2} (L/2 - x_0).
\end{aligned}
\label{eq:expression_w2}
\end{equation}
Dividing the rates in Eqs.\,(\ref{eq:expression_w1}) and (\ref{eq:expression_w2}) 
by $(L/2 - x_0)$, we get the expression for $P_{\text{imp}}$ as Eq.\,(\ref{eq:diss_rate_ls}).

Equations \,(\ref{eq:expression_w1}) and (\ref{eq:expression_w2}) indicate that the coupling between a ``chiral Fano'' system and the environment (the phonon bath) is enhanced by the on-resonance impurity.
Notice that Eqs.\,(\ref{eq:expression_w1}) and (\ref{eq:expression_w2}) are proportional to the length of the sample edge behind the impurity $L/2 - x_0$.
Since we have assumed negligible temperature variations on both $T_{\text{el}}$ and $T_{\text{lattice}}$, this result implies a constant energy dissipation rate per unit length (for the refinement, see the discussion around Fig.~\ref{fig:dissipation_rate} in the main text).

To further illustrate this, we consider $|M_s|^2$,
\begin{equation}
\begin{aligned}
&|M_s|^2 \! =\! \frac{1}{L^2} \frac{4 \Gamma^2}{ \epsilon_d^2 + \Gamma^2}  
\iint_{x_0}^{\frac{L}{2}}\! dx dx' e^{i(q_x - k_1 + k_2 )(x-x')} 
\\
& = \frac{1}{-i (q_x - k_1 + k_2 ) L^2} \frac{4 \Gamma^2}{ \epsilon_d^2 + \Gamma^2}  
\\
& \times \int_{x_0}^{\frac{L}{2}}\! dx 
[e^{-i(q_x - k_1 + k_2 )(\frac{L}{2}-x)}\! -\! e^{-i (q_x - k_1 + k_2 ) (x_0 - x)}],
\end{aligned}
\end{equation}
without taking the integral over $x$.
This way, the energy dissipation power becomes
\begin{equation}
\begin{aligned}
W_{\text{impurity}}& \ \!\propto \! \int_{x_0}^{\frac{L}{2}}\! dx \iint \! 
\frac{d^2\vec{q}}{i q_x}[ e^{i q_x (x-x_0)}\! - \! e^{iq_x(x- \frac{L}{2})}] \\
& \times \omega_q^3 |F(\vec{q})|^2 [ N_B^{\text{el}}(\omega_q) - N_B^{\text{lattice}} (\omega_q) ] \\
& = \int_{x_0}^{L/2} dx K(x),
\end{aligned}
\end{equation}
with the derivative of the kernel
\begin{equation}
\begin{aligned}
\frac{dK}{dx} &\! =\! \iint\! d^2\vec{q}\! 
\left\{ \cos[q_x(x-x_0)]\! -\! \cos[q_x ( x - L/2) ] \right\}\\
&\times \omega_q^3 |F(\vec{q})|^2[ N_B^{\text{el}}(\omega_q) - N_B^{\text{lattice}} (\omega_q) ].
\end{aligned}
\label{eq:deri}
\end{equation}
The major dissipation contribution in Eqs.\,(\ref{eq:expression_w1}) and (\ref{eq:expression_w2}) originates from the integral in the $q_y \sim q_T$ area, where Eq.\,(\ref{eq:deri}) vanishes. We have thus arrived at the conclusion that after the impurity position the impurity-induced energy dissipation rate per unit length should be a constant, if the temperature variation has been neglected.

\section{Dissipation contribution from $M_{\text{imp}}$}
\label{sec:imp_contribution}

As has been mentioned in Sec.\,\ref{sec:resonant scattering}, electrons may be trapped into the impurity, thus creating another matrix element $M_{\text{imp}}$.
In this section, we prove that $M_{\text{imp}}$ is negligible by calculating its vanishing contribution to the dissipation.

To begin with, under the presence of the impurity, the complete wave function contains three parts
\begin{equation}
\Phi_k(x,y) = \phi_k^{0} (x,y) +\phi_k^s(x,y) + \phi_k^d(x,y),
\label{eq:complete_wave_function}
\end{equation}
In Eq.\,(\ref{eq:complete_wave_function}), $\phi_k^{0} (x,y)$ is the free particle wave function. The second term $\phi_k^s(x,y)$ is the scattered state wave function that only exists at $x>0$.
The third term $\phi_k^d(x,y)$ is the part of the wave function that is trapped in the impurity.

Following Eq.\,(\ref{eq:complete_wave_function}), the overlap between the two wave functions $\iint dx dy \Phi^*_{k'}(x,y) \Phi_k(x,y)$, which equals $\delta_{k,k'}$ due to the orthogonality requirement, 
can be decomposed into three parts
\begin{subequations}
\begin{equation}
 \iint dx dy \phi_{k'}^{0*} \phi_k^0(x,y)
\label{eq:free_part}
\end{equation}
\begin{equation}
\begin{aligned}
&+ \iint dx dy [ \phi_{k'}^{s*} (x,y) \phi_k^s(x,y) \\
\ \ \ \ \ \ \ \ & + \phi_{k'}^{0*}(x,y) \phi_k^s(x,y) + \phi_{k'}^{s*}(x,y) \phi_k^{0}(x,y) ]
\end{aligned}
\label{eq:scatt_part}
\end{equation}
\begin{equation}
\begin{aligned}
+ & \iint dx dy[ \phi_{k'}^{d*}(x,y)\phi_k^d(x,y) \\
\ \ \ \ \ \ \ \ & + \phi_{k'}^{0*}(x,y)\phi_k^d(x,y) + \phi_{k'}^{d*}(x,y)\phi_k^0(x,y) \\
\ \ \ \ \ \ \ \ & + \phi_{k'}^{s*}(x,y)\phi_k^d(x,y) + \phi_{k'}^{d*}(x,y)\phi_k^s(x,y)].
\end{aligned}
\label{eq:imp_part}
\end{equation}
\label{eq:wave_function_overlap}
\end{subequations}

Because of the orthogonality of the first part, the sum of Eqs.\,(\ref{eq:scatt_part}) and (\ref{eq:imp_part}) equals zero.
Now we calculate the expression of $M_{\text{imp}}$ based on this equation.
With the definition $\vec{r} = (x,y)$ and the assumption that the point-like impurity is interacting with QH particles at a single point, $M_{\text{imp}}$ becomes
\begin{equation}
\begin{aligned}
M_{\text{imp}}
& \approx -e^{i\vec{q} \cdot \vec{r}_0} \iint dx dy ( \phi_{k'}^{s*} \phi_k^s + \phi_{k'}^{0*} \phi_k^s + \phi_{k'}^{s*} \phi_k^{0} ) \\
& = - e^{iq_x x_0} \frac{2i\Gamma v}{L(\epsilon_{k'} - \epsilon_d - i \Gamma) (\epsilon_k - \epsilon_d + i\Gamma)}\\
&\ \  \ \ \times [ e^{i(k-k') \frac{L}{2}} - e^{i(k-k') x_0} ] F(k,k'),
\end{aligned}
\end{equation}
where $\vec{r}_0 = (x_0,y_0)$ is the location of the impurity and the form factor in the $y$ direction
follows from the orthogonality requirement:
\begin{equation}
F(k,k') = \int dy \varphi^*_{k'}(y) \varphi_k(y) = \delta_{k,k'}.
\label{eq:form_factor_s}
\end{equation}
Consequently, $M_{\text{imp}}(k,k',\vec{q}) = 0$ unless $k = k'$, where no energy is dissipated.
We have thus arrived at the conclusion that $M_{\text{imp}}$ produces no dissipation once (i) the impurity is point-like and (ii) it only interacts with QH electrons at a single point.

In a more realistic consideration, we assume that the impurity has a finite size and interaction range $\sim a$. The form factor becomes $F(k,k',q_y) = \chi q_y a$ since $q_T a \ll 1$.
Following the same technique provided in Sec.\,\ref{sec:scatt_state_dissipation} we get the energy dissipation rate produced by $M_{\text{imp}}$
\begin{equation}
W_{\text{dot}} = \Gamma(5) \zeta(5) \frac{g_0^2}{(2\pi)^2} \frac{a^2 \xi^2}{ s^4} k_B^5[T_{\text{el}}^5 - T_{\text{lattice}}^5] \frac{4\Gamma^2}{\Gamma^2 + \epsilon_d^2}.
\label{eq:expression_wdot}
\end{equation}
A straightforward comparison between Eqs.\,(\ref{eq:expression_w2}) and (\ref{eq:expression_wdot}) shows that 
\begin{equation}
\frac{W_{\text{dot}}}{W_{\text{impurity}}} \sim \frac{a^2}{l_m L} \frac{v^2}{s^2} \frac{1}{q_T l_m} \ll 1.
\label{eq:comparison}
\end{equation}
Equation\,(\ref{eq:comparison}) proves that we can safely ignore $\phi_k^d$ from the wave function and the dissipation produced by $M_{\text{imp}}$.

\end{appendix}

\end{document}